\documentclass[11pt]{article}
\usepackage{moriond,epsfig}

\def\Journal#1#2#3#4{{#1} {\bf #2}, #3 (#4)}

\def\PLB{{\em Phys. Lett.}  B}
\def\PRL{\em Phys. Rev. Lett.}
\def\PRD{{\em Phys. Rev.} D}

\newcommand{\met}       {\mbox{$\not\!\!E_T$}}
\newcounter{N} 

\begin{document}
\vspace*{4cm}
\title{Top quark pair production cross section at Tevatron}

\author{V.~Shary for CDF and D0 collaborations}

\address{CEA, IRFU, SPP
Centre de Saclay, F-911191 Gif-sur-Yvette, France}

\maketitle\abstracts{
An overview of the recent measurements of the top antitop quark pair production cross section in proton antiproton 
collisions at $\sqrt{s}=1.96$ TeV in lepton + jets and  dilepton  final states is presented.
These measurements are based 
on  1 -- 2.8 fb$^{-1}$  of data collected with the D0 and CDF experiments at the Fermilab Tevatron collider. 
The cross section is measured with a precision close to 8 \% and found to be compatible with the standard model prediction.
Interpretations of the cross-section measurements for charge higgs search and for top 
quark mass measurement are also discussed.
}

\section{Introduction}
The measurements of the top antitop quark pair ($t\bar{t}$)
 production cross section are aimed to verify the agreement between 
the experimental data and the  perturbative  QCD calculation. 
Several approaches  for the $t\bar{t}$ cross section calculation to the next-to-leading order
are discussed in the literature \cite{shary:Nadolsky,shary:Cacciari,shary:Moch,shary:Kidonakis}. 
All of them have a comparable uncertainty between 7 \% and 10\%. Any significant deviation from the predicted value or
 differences in the measured cross section  in different final states may indicate  the presence of effects beyond standard model.

Within the standard model (SM) the top quark decays to a $W$ boson and $b$ quark with a probability close to 100\%. 
If both $W$ bosons (from top and antitop quarks) decay  to lepton-neutrino pairs, such final states will be referred to as
the ``dilepton'' channel. If one of the W bosons decays to a pair of light quarks, then this final state will be referred to as 
the ``lepton+jet'' channel. 
 
\section{Dilepton final state}

The decay signature of the dilepton final state consists of  two  leptons with large transverse momentum ($p_T$), two jets
originating from b~quarks and non-zero missing transverse energy (\met) due to the presence of 
neutrinos from W boson decays.
This channel has the highest signal to background ratio of  all  $t\bar{t}$ final states, but also the lowest probability,
$BR \sim 6.5\%$ for final states with $e$, $\mu$ or $\tau\to e,\mu$ and $BR \sim 3.6\%$ for final states with
hadronically decaying $\tau $ lepton.

The main sources of background in this channel are Z boson  ($Z/\gamma^\star\to l^+l^-$) and
diboson productions (WW, WZ and ZZ) with at least two charged leptons in the final state.
Other important background contribution is $W$ (+jets) and multijets production processes (so-called 
``instrumental background''). 
Semileptonic decays of b and c~quarks,  pion or kaon  decays could lead to an  additional muon in these processes.
An additional electron or tau lepton could be present because of the jet  misidentification.
$Z/\gamma^\star$ and diboson background contributions are estimated as 
$N_{bckg} = \sigma_{theory} \varepsilon\int L dt $, where $\sigma_{theory}$ is a theoretical cross section, 
$\int L dt$ is the integrated luminosity and $\varepsilon$ is the selection efficiency calculated using MC simulation.
Non of the existing MC generators can reproduce the $Z/\gamma^\star\ p_T$ 
distribution in data~\cite{Abazov:2008ez}. 
That is why this distribution is  reweighted according to the
data / MC difference measured in the  sample $Z/\gamma^\star\to l^+l^-$.
To estimate the instrumental background the CDF experiment assumes that the number of background events
in the main sample (where both leptons have  opposite charge) is equal to the number of events in the dilepton
samples where both  leptons have the same charge.  In addition to the ``same sign'' approach the D0 collaboration 
also estimates the instrumental background from a sample with inverted  quality on the lepton identification parameters.

The most recent CDF results are based on an integrated luminosity of 2.8 $fb^{-1}$ in three final states:
$ee$, $e\mu$ and $\mu\mu$. Two approaches are used to increase the signal-to-background ratio.
In the first  approach at least one jet is required to be identified as a jet originating from a b~quark  (``b-tagging'').
In the second approach a selection based purely on the kinematic properties of the final states is used.
The measured cross section is listed in  Table~\ref{shary:tab:CDFres}, 
results~(\ref{shary:res:cdf_dilep_kinem}) and (\ref{shary:res:cdf_dilep_btag}).
The most recent D0 results use an integrated luminosity of $\sim 1\ fb ^{-1}$  and include not only 
electron and muon final states,
but also the final states with hadronically decaying tau leptons:  $ee$, $e\mu$, $\mu\mu$, $e\tau$ and  $\mu\tau$.
The measured  cross section  is  listed in Table~\ref{shary:tab:D0res}, result~(\ref{shary:res:d0_dilep}).
The  cross section measurements in final states with hadronically decaying tau leptons are important to constrain 
 non standard model contribution to the $t\bar{t}$ final states (see an example in section~\ref{shary:sec:intr}).
The D0 2.2~$fb^{-1}$ measurement in the $e\tau$ and  $\mu\tau$ final states gives a value 
shown in Table~\ref{shary:tab:D0res}, result~\ref{shary:res:d0_taulep}.

 %======================================================

\begin{table}
\rule{\textwidth}{1pt}
\begin{list}{(\arabic{N})}{\usecounter{N}\itemsep=0cm\parsep=1pt\topsep=0pt\parskip=0pt}
\item dileptons, kinematic based approach (2.8 $fb^{-1}$) \cite{shary:CDF:dilepton} : \hfill
$\sigma_{t\bar{t}} = 6.7 \pm 0.8  \pm 0.4  \pm 0.4 \  pb$ 
\label{shary:res:cdf_dilep_kinem} 

\item dileptons, b-tagging  approach (2.8 $fb^{-1}$) \cite{shary:CDF:dilepton} : \hfill
 $\sigma_{t\bar{t}} = 7.8 \pm 0.9  \pm 0.7  \pm 0.45 \ pb$
\label{shary:res:cdf_dilep_btag} 

\item lepton+jets, kinematic based approach  (2.8 $fb^{-1}$) \cite{shary:CDF:ljets_NN} :  \hfill
$\sigma_{t\bar{t}} = 7.1 \pm 0.4  \pm 0.4  \pm 0.4 \  pb$
\label{shary:res:cdf_lepjets_kinem} 

\item lepton+jets, b-tagging  (2.7 $fb^{-1}$) \cite{shary:CDF:ljets_btag}:  \hfill
$\sigma_{t\bar{t}} = 7.2 \pm 0.4  \pm 0.5  \pm 0.4 \ pb$
\label{shary:res:cdf_lepjets_btag} 

%\item lepton+jets, b-tagging with soft muon (2.0 $fb^{-1}$) \cite{Aaltonen:2009ax} : \hfill
%$\sigma_{t\bar{t}} = 9.1 \pm 1.1  ^{+1.0}_{-0.9}  \pm 0.6  pb $

%\item lepton+jets, b-tagging with soft electron (2.0 $fb^{-1}$) \cite{shary:CDF:ljets_etag} : \hfill
%$\sigma_{t\bar{t}} = 7.8 \pm 2.4  \pm 1.5  \pm 0.5  pb$

\item alljets (1.0 $fb^{-1}$) \cite{Aaltonen:2007qf} : \hfill
$\sigma_{t\bar{t}} =  8.3 \pm 1.0  ^{+2.0}_{-1.5}  \pm 0.5 \ pb$

\item[] \textbf{CDF combined, preliminary ($\mathbf{2.8\ fb^{-1}}$) \cite {shary:CDF:comb}} : \hfill 
$\sigma_{t\bar{t}} = 7.0\pm 0.6 \ pb$
\end{list}
\rule{\textwidth}{1pt}
\caption{\label{shary:tab:CDFres}$t\bar{t}$ cross section measurements by the CDF experiment. The first uncertainty 
is the statistical only, the second one is the systematic uncertainty and the last one is the uncertainty on 
the integrated luminosity.  For the combined result all 
uncertainties are combined together.
 In all measurements $m_t$ is assumed to be 175 GeV.}
\end{table}

%============================================

\begin{table}
\rule{\textwidth}{1pt}
\begin{list}{(\arabic{N})}{\usecounter{N}\itemsep=0cm\parsep=1pt}
\item dileptons (1.0 $fb^{-1}$) \cite{Abazov:2009si} : \hfill
$\sigma_{t\bar{t}} = 7.5 \pm  1.0  ^{+0.7}_{-0.6}  \pm 0.6 \ pb$  ($m_t = 170$~GeV)
\label{shary:res:d0_dilep} 

\item $\tau$+lepton	 (2.2 $fb^{-1}$) \cite{shary:D0:taulepton}   : \hfill
$\sigma_{t\bar{t}} = 7.3 ^{+1.3}_{-1.2}\ ^{+1.2}_{-1.1}  \pm 0.45  \ pb$ ($m_t = 170$~GeV) 
\label{shary:res:d0_taulep} 

\item lepton+jets (0.9 $fb^{-1}$) \cite{Abazov:2008gc} : \hfill
$\sigma_{t\bar{t}} = 7.8 \pm 0.5  \pm 0.5  \pm 0.45 \  pb$ ($m_t = 175$~GeV) 
\label{shary:res:d0_lepjets} 

\item alljets (0.4 $fb^{-1}$) \cite{Abazov:2006yb} : \hfill
$\sigma_{t\bar{t}} = 4.5^{+2.0}_{-1.9}\ ^{+1.4}_{-1.1}  \pm  0.3 \ pb$
($m_t = 175$~GeV)

\item[] \textbf{D0 combined, preliminary ($\mathbf{1.0\ fb^{-1}}$) \cite{Abazov:2009ae}}
: \hfill $\sigma_{t\bar{t}} = 8.2^{+1.0}_{-0.9}\ pb$ ($m_t = 170$~GeV)
\end{list}
\rule{\textwidth}{1pt}
\caption{\label{shary:tab:D0res}$t\bar{t}$ cross section measurements by the D0 experiment. The first uncertainty 
is the statistical only, the second one is the systematic uncertainty and 
the last one is the uncertainty on the integrated luminosity. For the combined result all 
uncertainties are combined together.}
\end{table}

%===========================================================

\section{Lepton + jets final state}
The decay signature for the lepton+jets final state consists of  one  high $p_T$ lepton, two b~quark jets,
two jets originating from the W boson decay  and non-zero \met\  due to the presence 
of the neutrino from the leptonic decay of the second W boson.
This channel has high branching ratio, $BR \sim 35\%$ for final state including 
$e$, $\mu$, $\tau\to e,\mu$ 
and $BR \sim 9.5\%$ for final states containing a hadronically decaying $\tau $ lepton.
At the same time the background contribution in this final state is higher than in the dileptonic one. The main 
sources of background are W+jets and multijet production processes. 
In addition to the selection which  require one high $p_T$ lepton, at least three jets and high \met, 
both experiments  use two approaches to reduce the background contribution. 
The first one is based on the b~quark jet identification and 
the second approach uses a multivariate discriminant built with  kinematic information only.
The multijet background is determined from a sample enhanced with multijet events by loosening the lepton identification 
criteria. The W+jets background distributions shapes are determined from MC simulation, 
but the overall normalization is adjusted  to data.
For W+jets simulation both experiments use the combination
of matrix element generator Alpgen~\cite{Mangano:2002ea} 
with showering generator Pythia~\cite{shary:pythia}.
The heavy flavor contributions are adjusted 
by scaling up corresponding cross sections with scale factors determined from data.
The $t\bar{t}$  cross sections measured by the CDF experiment with 2.8~$fb^{-1}$ are shown in 
Table~\ref{shary:tab:CDFres}, results~(\ref{shary:res:cdf_lepjets_kinem}) and (\ref{shary:res:cdf_lepjets_btag}). 
The  D0 0.9 $fb^{-1}$ measurement  which combine both b-tagging and kinematic based approaches 
is listed in Table~\ref{shary:tab:D0res}, result~(\ref{shary:res:d0_lepjets}).

The b-tagging approach  provides a purer sample of $t\bar{t}$ events than the one using only kinematic information,
 but the systematic uncertainty of the cross section measurement is slightly higher, which is explained by the additional 
uncertainty due to the b~quark identification procedure  (5\% for the CDF and 6\% for D0).  A better understanding 
of the b-tagging procedure may improve this uncertainty, but 
both approaches (the b-tagging  and kinematic ones) will stay  limited 
by the uncertainty on the integrated luminosity measurement , 
5.8\% for CDF, 6.1\% for D0 (see complete systematic breakdown tables in 
\cite{shary:CDF:ljets_NN,shary:CDF:ljets_btag,Abazov:2008gc}).
 That is why the CDF collaboration explores a new way of determining the 
$t\bar{t}$ cross section by measuring a ratio of $t\bar{t}$ to Z boson cross sections.
This ratio is insensitive to the integrated luminosity uncertainty and ``replaces'' it with an
uncertainty on the calculated value of the Z boson cross section.
Using the Z boson cross section in the invariant mass range 66 -- 116 GeV 
($\sigma_Z = 251.3 \pm 5.0\ pb$~\cite{Abulencia:2005ix})
the measured $t\bar{t}$ cross section is found to be \cite{shary:CDF:ljets_NN,shary:CDF:ljets_btag}: 
\begin{itemize}
\item[-] kinematic based analysis: 
$\sigma_{t\bar{t}} = 6.9 \pm 0.4 (stat) \pm 0.6 (syst) \pm 0.1 (theory)\ pb$, $m_t = 175$~GeV 
\item[-] b-tagging analysis:  \hfill
$\sigma_{t\bar{t}} = 7.0 \pm 0.4 (stat)  \pm 0.4 (syst) \pm 0.1 (theory)\ pb$, $m_t = 175$~GeV
\end{itemize}

%==================================================

\section{Interpretations of $t\bar{t}$ cross section measurements}
\label{shary:sec:intr}

\begin{figure}
\begin{minipage}[t]{.61\textwidth}
\includegraphics[width=.495\textwidth]{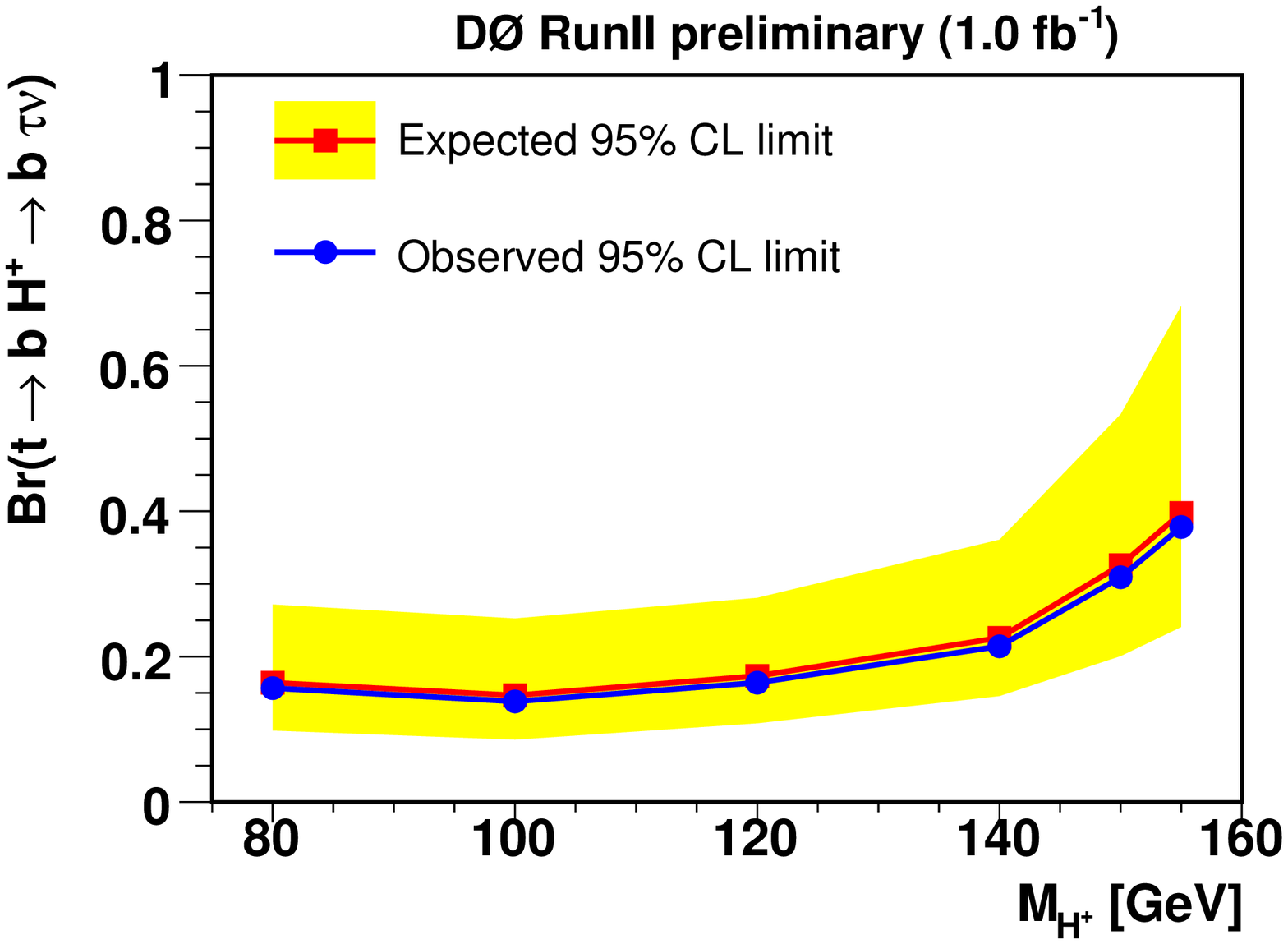}
\hfill
\includegraphics[width=.495\textwidth]{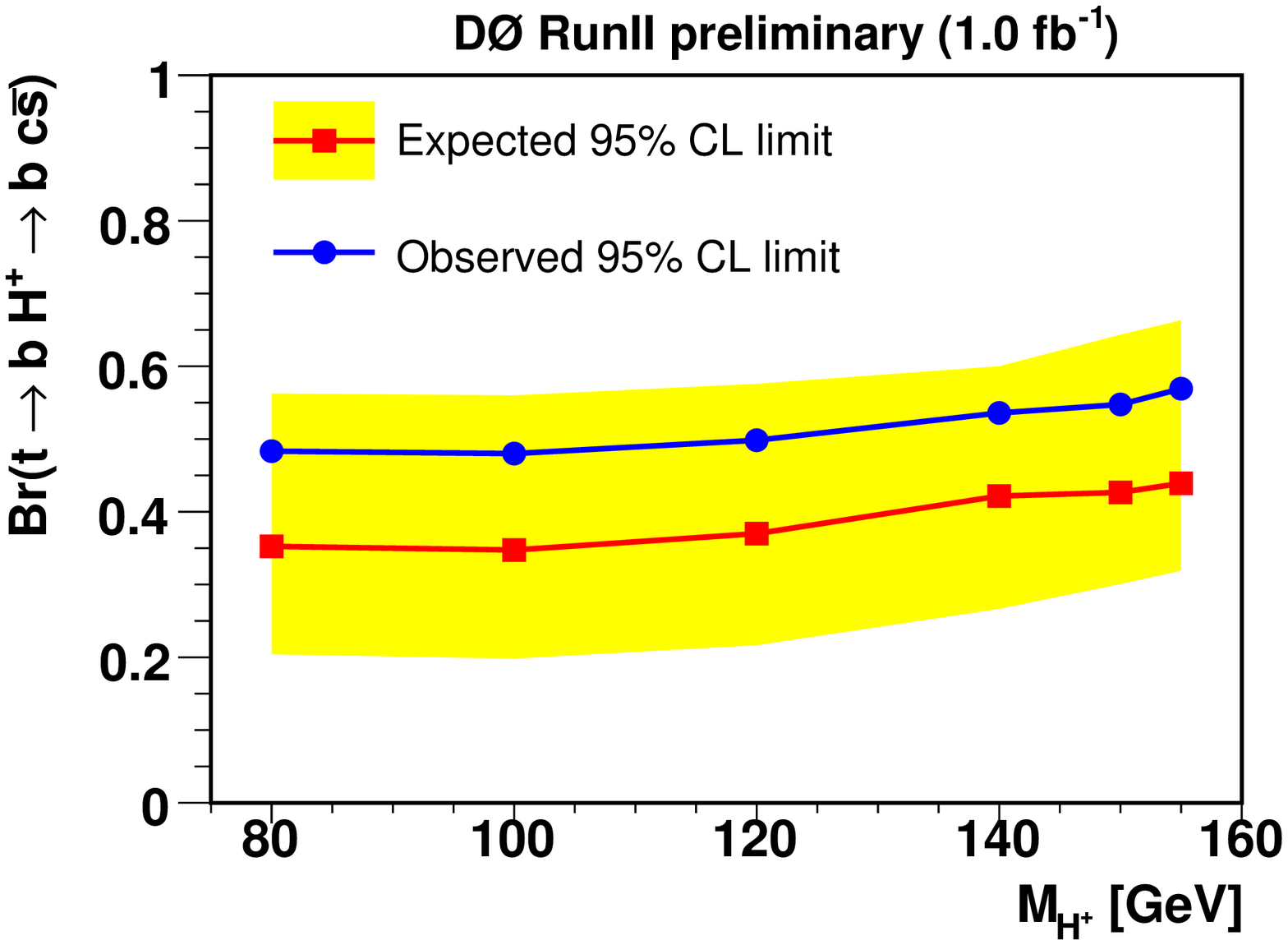}
\caption{Upper limits on $B(t \to H^+b)$ for tauonic (left) and leptophobic (right) 
$H^+$ decays. The yellow band shows the $\pm 1$ standard deviation band 
around the expected limit.
\label{shary:fig:limits}}
\end{minipage}
\hfill
\begin{minipage}[t]{.35\textwidth}
\includegraphics[width=\textwidth]{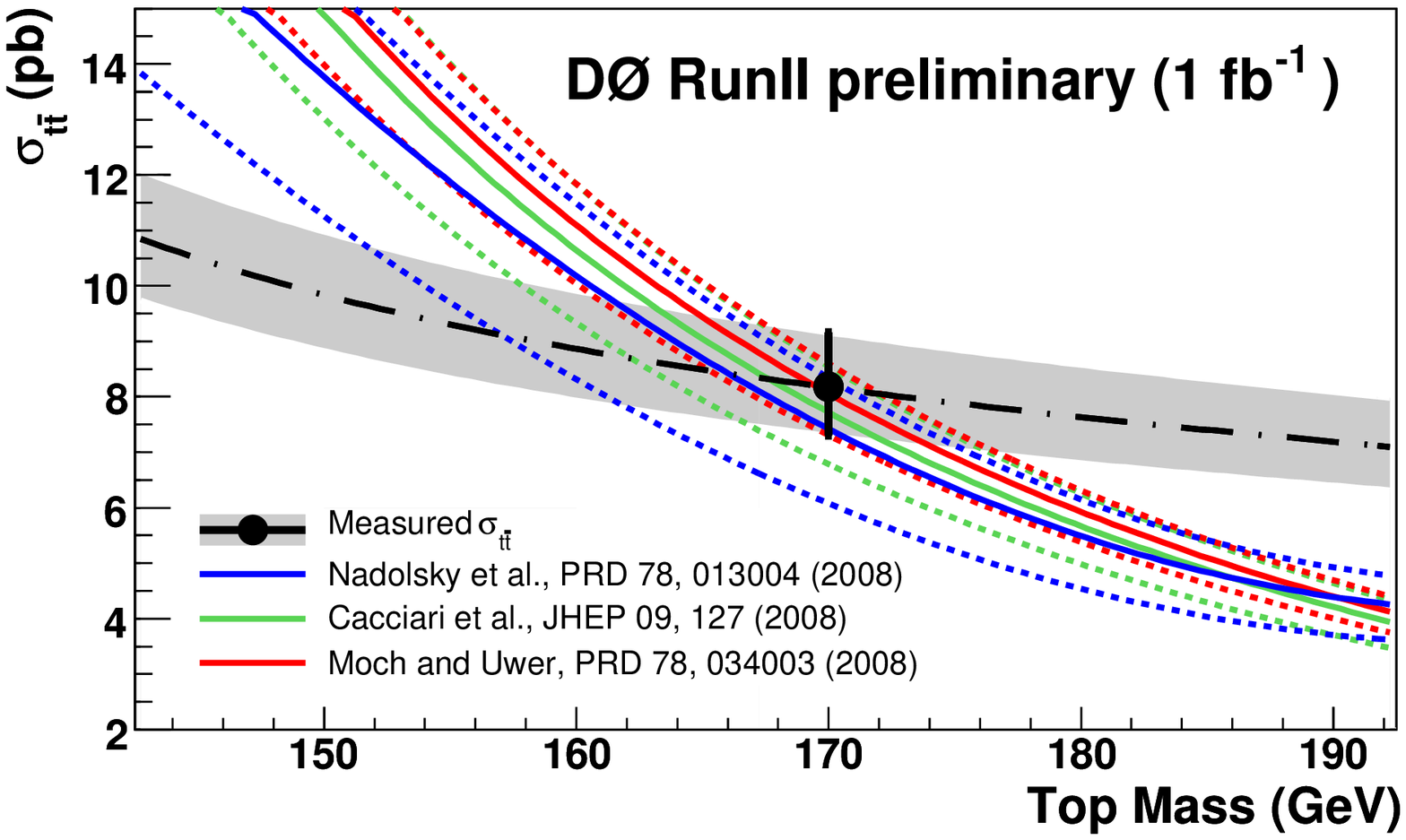}
\caption{Experimental (point and black dashed line) and theoretical $\sigma_{t\bar{t}}$ as a function of $m_t$. 
The gray band is the total experimental uncertainty.
\label{shary:fig:mass}}
\end{minipage}
\end{figure}

The ratio of cross sections measured in different final states are particularly sensitive to new physics which may appear
in top quark decays, especially if the boson from the top decay is not a W boson.
For example, the decay into a charged Higgs boson ($t \to H^+b$) as 
predicted in some models~\cite{shary:chargehiggs}, can compete with the SM decay $t \to W^+b$.
Using the ratios $\sigma(t\bar{t})_{dilepton} / \sigma(t\bar{t})_{lepton+jets}$ and 
$\sigma(t\bar{t})_{\tau+lepton} / \sigma(t\bar{t})_{dilepton\ \&\ lepton+jets}$
the D0 experiment extracts an upper limit on the branching ratio $B(t\to H^+b)$ in case of the leptophobic
($H^+\to c\bar{s}$) and tauonic ($H^+\to \tau\nu$) models respectively~\cite{Abazov:2009ae}.
The corresponding limits are shown in Fig.~\ref{shary:fig:limits}.

Another interesting interpretation of the $t\bar{t}$ cross section measurement is the extraction of the top quark mass
using the theoretical dependence which relates cross-section with mass.
This provides a  measurement complementary to the direct top quark mass measurement, which is done 
in a well defined renormalization scheme,  employed in the theoretical cross section calculation.
Fig.~\ref{shary:fig:mass} shows the D0 combined experimental and the theoretical 
\cite{shary:Nadolsky,shary:Cacciari,shary:Moch}
cross sections as a function of the top quark mass.  Following the
method in~\cite{Abazov:2009si,Abazov:2008gc}, the D0 collaboration
extracts the top quark mass value at  68\% CL. Since the theoretical
calculations are performed in the pole mass scheme, this defines the
extracted parameter here.  The results  \cite{Abazov:2009ae} are given in
Table~\ref{shary:tab:mass}. All values are in good agreement with the
current world average of $173.1\pm 1.3$~GeV~\cite{topmass:2009ec}.

\begin{table}
\begin{tabular}{l|r}
\hline \hline
Theoretical computation & $m_t$ (GeV) \\ \hline
NLO \cite{shary:Nadolsky} & $165.5^{+6.1}_{-5.9}$ \\ \hline
NLO+NLL \cite{shary:Cacciari} & $167.5^{+5.8}_{-5.6}$\\ \hline
\hline
\end{tabular}
\hfill
\begin{tabular}{l|r}
\hline \hline
Theoretical computation & $m_t$ (GeV) \\ \hline
approximate NNLO \cite{shary:Moch} &  $169.1^{+5.9}_{-5.2}$\\ \hline
approximate NNLO \cite{shary:Kidonakis} & $168.2^{+5.9}_{-5.4}$\\ \hline
\hline
\end{tabular}
\caption{\label{shary:tab:mass} Top quark mass at 68\% C.L. for different theoretical computations
 of the $t\bar{t}$ cross section. Combined experimental and theoretical uncertainties are shown.}
\end{table}

%=========================================================

\end{document}